\documentclass[twocolumn,pre,twoside,floatfix]{revtex4-1}

\usepackage{amsmath}
\usepackage{amssymb}
\usepackage{graphicx}
\usepackage{dcolumn}
\usepackage{bm}
\usepackage{color}
\usepackage[T1]{fontenc}

\def\dps{\displaystyle}

\def\d{\mathrm{d}}

\def\epsilon{\varepsilon}
\def\erfc{\mathop\mathrm{erfc}}
\def\theta{\vartheta}
\def\rho{\varrho}

\def\vec#1{\mathbf{#1}}

\graphicspath{{Figs/}}

\begin{document}


\title{Thermal and structural properties of ionic fluids}

\author{Hendrik Bartsch}
\email{hbartsch@is.mpg.de}
\author{Oliver Dannenmann}
\author{Markus Bier}
\email{bier@is.mpg.de}
\affiliation
{
   Max-Planck-Institut f\"ur Intelligente Systeme, 
   Heisenbergstr.\ 3,
   70569 Stuttgart,
   Germany, 
   and
   Institut f\"ur Theoretische Physik IV,
   Universit\"at Stuttgart,
   Pfaffenwaldring 57,
   70569 Stuttgart,
   Germany
}

\date{15 April 2015}

\begin{abstract}
The electrostatic interaction in ionic fluids is well-known
to give rise to a characteristic phase behavior and structure. 
Sometimes its long range is proposed to single out the electrostatic potential
over other interactions with shorter ranges.
Here the importance of the range for the phase behavior and the structure of
ionic fluids is investigated by means of grandcanonical Monte Carlo simulations
of the lattice restricted primitive model (LRPM).
The long-ranged electrostatic interaction is compared to various types of
short-ranged potentials obtained by sharp and/or smooth cut-off schemes.
Sharply cut off electrostatic potentials are found to lead to a strong 
dependence of the phase behavior and the structure on the cut-off radius.
However, when combined with a suitable additional smooth cut-off, the 
short-ranged LRPM is found to exhibit quantitatively the same phase behavior
and structure as the conventional long-ranged LRPM.
Moreover, the Stillinger-Lovett perfect screening property, which is well-known
to be generated by the long-ranged electrostatic potential, is also fulfilled
by short-ranged LRPMs with smooth cut-offs.
By showing that the characteristic phase behavior and structure of ionic fluids
can also be found in systems with short-ranged potentials, one can conclude 
that the decisive property of the electrostatic potential in ionic fluids is
not the long range but rather the valency dependence.
\end{abstract}

\maketitle


\section{Introduction}

Over the last two decades, the scientific interest in organic salts with 
melting points near room temperature, so-called room temperature ionic liquids
(RTILs), has been growing steadily. 
Features such as a negligible vapor pressure and a remarkable
thermal stability \cite{Liu_et_al,Paulechka_et_al,Bermudez_et_al,Bier2010}
promise future applications as solvents for syntheses, electrolytes in fuel
cells, solar cells, and batteries, as well as in biomass processing 
\cite{Yamanaka_et_al,Wasserscheid2007,Greaves_Drummond,Armand_et_al,
Tan_MacFarlane2010,Yasuda_Watanabe}.
Moreover, due to the tiny vapor pressure many applications are
conceivable under ultrahigh vacuum \cite{Liu_et_al,Wang_et_al,Suzuki_et_al,
Bermudez_et_al}.
However, the technological use of RTILs requires an in-depth understanding of
ionic systems.

The Coulomb interaction, underlying these systems, acts repulsively for 
equally-charged and attractively for oppositely-charged ions and it decays 
$\propto1/r$, i.e., it is long-ranged.
On the other hand, the pair distribution functions of an ionic fluid decay 
exponentially, which is called the Stillinger-Lovett perfect screening property
\cite{Stillinger_Lovett,Lovett_Stillinger,Stillinger_Lovett2,Hansen2006}
and which is a necessary consequence of the long range of the Coulomb potential
\cite{Mitchell_et_al,Martin_Gruber}.
It is a challenge for more than a century now to develop 
quantitatively reliable theoretical descriptions of this peculiar
combination of properties.
For dilute electrolyte solutions Debye-H\"{u}ckel theory \cite{1Fisher,
Lee_Fisher,Kobelev_et_al} is typically a good starting point, whereas for 
dense ionic systems ion-pairing is relevant \cite{Stillinger_Lovett,
Lovett_Stillinger}. 
From experimental work \cite{1Wiegand_et_al,3Wiegand_et_al,2Wiegand_et_al,
Bianchi_et_al} as well as continuum simulations \cite{Yan_dePablo,
Luijten_et_al,1Panagiotopoulos,Kim_Fisher}, there is evidence that in 
particular the critical behavior of ionic systems is very similar to the Ising
universality class, which typically applies to systems with
short-ranged interactions.
This result suggests that the long-range character of the bare
Coulomb potential might be of minor importance for the critical
properties of ionic fluids.

The aim of the present work is to investigate the relevance of the
long range of the Coulomb interaction for the whole phase diagram and the 
bulk structure.
This is done by studying the lattice restricted primitive model (LRPM) for the
Coulomb potential being truncated smoothly on a length
scale $1/\alpha$ and sharply at a cut-off radius $r_\text{cut}$.
The technical details of the model and the approach are presented
in Sec.~\ref{sec:theory}.
The long-ranged LRPM is well-known to exhibit tricritical behavior
at the crossover from a first- to a second-order phase transition 
between a charge-ordered and a charge-disordered phase \cite{Dickman_Stell,
Ren_et_al}.
The (L)RPM renders the ions as homogeneous charged hard spheres,
which is of course not valid for ionic liquids in general,
since they may exhibit charge as well as shape anisotropy.
However, this work is concerned with the influence of the long range
and the valency dependence of the electrostatic interaction,
being omnipresent in all kinds of ionic systems.
In particular, the impact of the long range will not depend
on rather short-range features like charge or shape anisotropy,
therefore we consider this simple model of
ionic fluids as sufficient for the present investigation.
It is shown in Sec.~\ref{sec:discussion} that, upon varying the 
smooth cut-off decay constant $\alpha$ and the sharp cut-off radius 
$r_\text{cut}$, the short-ranged LRPM can be tuned from a system void of a
charged-ordered phase via one exhibiting charge-ordered and 
charge-disordered phases but differing quantitatively from the long-ranged 
LRPM to a model with short-ranged interactions whose phase behavior and 
structure is quantitatively the same as for the long-ranged Coulomb 
interaction.
Moreover, even the Stillinger-Lovett perfect screening property can be 
fulfilled for suitable smooth cut-off potentials, an observation which is
not trivial in the context of short-ranged interactions. 
By showing that the characteristic phase behavior and structure of ionic fluids
can also be found in systems with short-ranged potentials, it is concluded
in Sec.~\ref{sec:conclusions} that the decisive property of the electrostatic
potential in ionic fluids is not the long range but rather the valency 
dependence.


\section{\label{sec:theory}Model and method}

\subsection{\label{sec:theory:model}LRPM with truncated Coulomb potential}

Consider the lattice restricted primitive model (LRPM) of univalent cations 
(valency $z_\oplus=+1$) and anions (valency $z_\ominus=-1$) with hard cores of
diameter $\sigma$ occupying but not necessarily exhausting the sites of a 
three-dimensional simple cubic lattice with lattice constant $\sigma$.
Global charge neutrality is guaranteed by an equal number of cations and anions
in the system, and the hard cores ensure that each site is either empty or 
singly occupied.
Within the original LRPM two ions of species $i,j\in\{\oplus,\ominus\}$ at a
distance $r$ interact, besides the hard core exclusion, via the 
infinitely-ranged Coulomb potential $\dps \frac{l_bz_iz_j}{r}$, where $\dps 
l_b=\frac{e^2}{4\pi\epsilon kT}$ is the Bjerrum length with the electronic 
permittivity $\epsilon$, the Boltzmann constant $k$, and the temperature $T$,
and where $r$ is measured with the Euclidean metric.

In the present work the relevance of the long-range character of the Coulomb
interaction for the properties of ionic systems is studied by considering the
implications of replacing the infinitely-ranged Coulomb potential by the
truncated Coulomb-like potential
\begin{equation}
   \beta\phi_{p,q}(\vec{r},z_i,z_j) = 
   \begin{cases}
      \dps\frac{l_bz_iz_j}{\|\vec{r}\|_p}\erfc(\alpha\|\vec{r}\|_p) 
      & , \|\vec{r}\|_q \leq r_\text{cut} \\
      0 
      & , \|\vec{r}\|_q > r_\text{cut}
   \end{cases}
\label{eq:phi_ij}
\end{equation}
with the inverse temperature $\beta=1/(kT)$, the decay 
constant $\alpha\geq0$, and the cut-off radius $r_\text{cut}$.
By means of Eq.~(\ref{eq:phi_ij}) two different methods of cutting 
off the Coulomb potential can be studied: On the one hand, the factor 
$\erfc(\alpha\|\vec{r}\|_p)$,
leads to a smooth cutting off on the length scale
$1/\alpha$, and, on the other hand, a sharp cutting off at radius 
$\|\vec{r}\|_q=r_\text{cut}$ can be considered.
Inspired by the short-ranged potentials appearing within the Ewald method~\cite{Ewald},
here the smoothening factors are chosen to be complementary error functions
$\erfc(x)=2/\pi\int_x^\infty\d\tau\exp\left(-\tau^2\right)$.
Obviously, the infinitely-ranged Coulomb potential corresponds to 
$\alpha=0, r_\text{cut}=\infty$.
Besides the Euclidean norm $\|\vec{r}\|_2=\sqrt{x^2+y^2+z^2}$ to measure 
distances, the $1$-norm $\|\vec{r}\|_1=|x|+|y|+|z|$ and the supremum norm 
$\|\vec{r}\|_\infty=\max(|x|,|y|,|z|)$ are considered, which are more 
adapted to a lattice model since, for lattice vectors $\vec{r}$, they lead to
values which are integer multiples of the lattice constant $\sigma$.
Since all norms are equivalent in finite dimensions, the power law $\propto 
1/r$ is asymptotically preserved irrespective of the choice of the norm. 
The parameters $p$ and $q$ in Eq.~(\ref{eq:phi_ij}) describe the
norms to be used for measuring the distance determining the interaction
potential and the sharp cut-off, respectively.


\subsection{\label{sec:theory:mc}Grandcanonical Monte Carlo simulations}

In order to discuss the thermal and structural properties of the LRPM with the
truncated Coulomb-like interaction Eq.~(\ref{eq:phi_ij}), the packing fraction
$\eta$ and the pair distribution functions $g_{ij}(r)$, 
$i,j\in\{\oplus,\ominus\}$, are determined for cubic boxes $\mathcal{V}:=
\{0,\sigma,\dots, (L-1)\sigma\}^3, L\in\mathbb{N},$ with periodic boundary 
conditions using grandcanonical Monte Carlo simulations.
The set of all configurations $\zeta$ of cations and anions occupying 
$\mathcal{V}$ can be expressed as the set of all maps $\zeta:\mathcal{V}\to
\{0,z_\oplus,z_\ominus\}$, which result in the charge $\zeta(\vec{r})\in
\{0,z_\oplus,z_\ominus\}$ located at position $\vec{r}\in\mathcal{V}$, i.e., 
$\zeta(\vec{r})=0$ iff site $\vec{r}$ is empty, $\zeta(\vec{r})=z_\oplus$ iff
site $\vec{r}$ is occupied by a cation, and $\zeta(\vec{r})=z_\ominus$ iff site
$\vec{r}$ is occupied by an anion.
Standard Metropolis importance sampling of the grandcanonical Boltzmann 
distribution $\dps P(\zeta) \propto \exp(\beta\mu N[\zeta] - \beta H[\zeta])$
with the chemical potential $\mu$, the total number of ions $N[\zeta]=
N_{\oplus}[\zeta]+N_{\ominus}[\zeta]$, and the Hamiltonian
\begin{align}
   \beta H[\zeta] = \frac{1}{2}
   \sum_{\substack{\vec{r},\vec{r'}\in\mathcal{V}\\\vec{r}\not=\vec{r'}}}
   \beta\phi_{p,q}\left(\vec{r}-\vec{r'},\zeta(\vec{r}),\zeta(\vec{r'})\right)
   \label{eq:ham}
\end{align}
on the set of all charge-neutral configurations $\zeta$ is applied.
Charge neutrality is preserved during the simulation runs due to insertions and
removals of only neutral pairs of cations and anions, i.e., the number of 
cations $N_\oplus[\zeta]$ equals the number of anions $N_\ominus[\zeta]$.

The packing fraction is the average
\begin{align}
   \eta = \frac{\langle N[\zeta] \rangle}{L^3}
\end{align}
and the pair distribution functions are given by
\begin{equation}
   g_{i,j}(r) = 
   \frac{4\langle N_{i,j}(r,[\zeta])\rangle}{V_\text{shell}(r)L^3\eta^2}
   \label{eq:pdf} 
\end{equation}
with $i,j\in\{\oplus,\ominus\}$, $r$ being a distance measured in the $1$-norm,
$V_\text{shell}(r)=4(r/\sigma)^2+2$ representing the number of all sites in the
shell of $1$-norm distance $r$ around a site, and $N_{i,j}(r,[\zeta])$ denoting
the total number of all ordered pairs $(\vec{r}_i,\vec{r}_j)$ of positions 
$\vec{r}_i,\vec{r}_j\in\mathcal{V}$ being separated by a $1$-norm distance 
$r=\|\vec{r}_i-\vec{r}_j\|_1$ and such that an ion of species $i$ is located at
position $\vec{r}_i$ and an ion of species $j$ is located at position 
$\vec{r}_j$.
Here the pair distribution functions $g_{i,j}(r)$ are defined in terms of
$1$-norm distances $r$ because, for the present simple-cubic lattice geometry,
this choice is most convenient in order to distinguish charge-ordered and 
charge-disordered phases.

The equilibrium state of the LRPM is determined by the choice of the 
(dimensionless) temperature $T^*:=\sigma/l_b$ and the (dimensionless) chemical
potential $\mu^*:=\beta\mu$.
Within the present work we considered temperatures and chemical 
potentials in the ranges $T^*\in[0.04,10]$ and $\mu^*\in[-8,0]$, 
respectively.
Box sizes were varied in the range $L\in\{8,\dots,19\}$.

From an analysis of the equation of state, which are investigated as isotherms
in $\eta$-$\mu^*$-space, and of the structure in terms of the pair distribution
functions the phase diagram of the LRPM for various sets of values
of the decay constant $\alpha$ and of the cut-off radius $r_\text{cut}$ 
are inferred.
The following subsections explain the procedures to estimate the locations of
the binodals of the first-order gas/liquid phase transition and of the 
so-called $\lambda$-line of the continuous phase transition between the 
charge-ordered and the charge-disordered phase.


\subsection{\label{sec:theory:binodal}Estimation of the binodals}

\begin{figure}[!t]
   \includegraphics{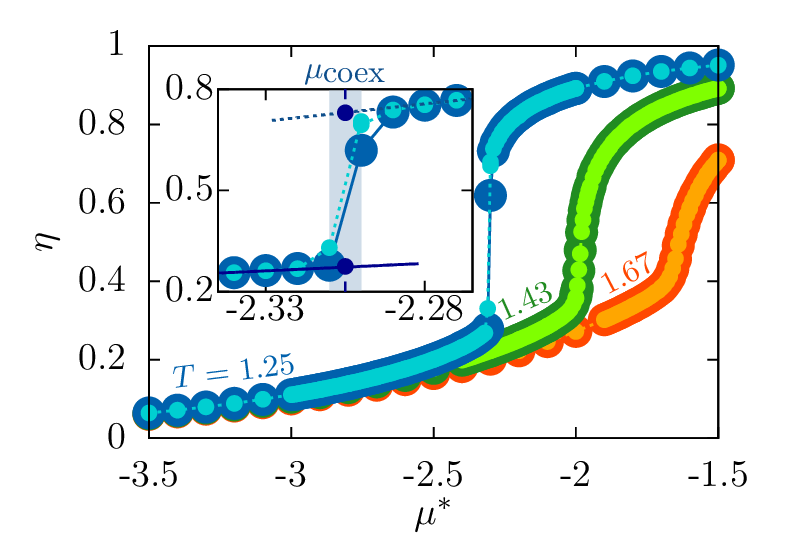}
   \caption{
   Isotherms $T^*\in\{1.25,1.43,1.67\}$ for $L=12,\alpha=0,
   r_\text{cut}/\sigma=1$ using the $1$-norm ($p=q=1$) in Eq.~(\ref{eq:ham}).
   For $T^*=1.25$ a distinct discontinuity of the packing fraction $\eta$ at
   chemical potential $\mu^*\approx-2.3$ can be observed, while for $T^*=1.67$
   a continuous curve occurs. 
   Therefore, the critical temperature $T^*_c$, where the first-order 
   gas/liquid phase transition associated with the discontinuity of the packing
   fraction $\eta$ terminates, is located within this temperature range: 
   $T^*_c\in[1.25,1.67]$. 
   For each isotherm two sets of data points are shown, which correspond to
   either an initially empty system (large dots connected by solid lines) or an
   initially totally filled system (small dots connected by dashed lines). 
   The inset sketches the procedure to estimate the packing fractions at 
   coexistence (see Sec.~\ref{sec:theory:binodal}).
   }
   \label{fig:isotherms_r1}
\end{figure}

Estimates of the binodals of the first-order gas/liquid phase transition are
obtained by determining the equation of state $\eta(T^*,\mu^*)$.
Figure~\ref{fig:isotherms_r1} displays isotherms $T^*\in\{1.25,1.43,
1.67\}$ for $L=12, \alpha=0, r_\text{cut}/\sigma=1$ using the 
$1$-norm ($p=q=1$) in Eq.~(\ref{eq:ham}).
For the lowest temperature $T^*=1.25$ a discontinuity in the packing fraction 
$\eta(\mu^*)$ indicates a first-order gas/liquid phase transition, whereas for
the highest temperature $T^*=1.67$ no gas/liquid phase transition occurs.
An estimate of the critical temperature $T^*_c$ is obtained as a temperature
for which $\eta(\mu^*)$ is discontinuous for $T^*<T^*_c$ and continuous for
$T^*>T^*_c$.
This simple method is sufficiently precise for the present purpose.

In order to estimate the binodal packing fractions $\eta_1(T^*)$ and 
$\eta_2(T^*)$ of gas/liquid coexistence at a temperature $T^*<T^*_c$ the phase
transition is first located within a range of the chemical potentials $\mu^*$
which is as narrow as possible and outside which the well-known hysteresis 
effect of first-order phase transitions does not occur.
For $T^*=1.25$ in Fig.~\ref{fig:isotherms_r1}, e.g., the inset leads to the 
range $\mu^*\in[-2.33,-2.28]$. 
Our estimate of the chemical potential $\mu^*_\text{coex}(T^*)$ at coexistence
is the midpoint of the $\mu^*$-interval of steepest increase of the packing
fraction $\eta(\mu^*)$.
For the isotherm depicted in the inset of Fig.~\ref{fig:isotherms_r1}, this 
choice yields $\mu^*_\text{coex}(T^*)\approx-2.305$ for $T^*=1.25$.
Estimates of the binodal packing fractions $\eta_1(T^*)$ and $\eta_2(T^*)$
are obtained by linearly extrapolating the hysteresis-free parts of 
$\eta(\mu^*)$ to $\mu^*=\mu^*_\text{coex}(T^*)$ from below and from above,
respectively.
For Fig.~\ref{fig:isotherms_r1}, e.g., these estimates are $\eta_1(T^*)\approx
0.27$ and $\eta_2(T^*)\approx0.73$ for $T^*=1.25$.
Again, this simple method is sufficiently precise,
since, for the purpose of the present work, the location of the binodals is
required to be known only semi-quantitatively but not to the highest technically
feasible accuracy.
More sophisticated and thus precise methods,
such as umbrella sampling techniques
or methods based on finite-size scaling and histogram reweighting
\cite{Privman1990,Ferrenberg_Swendsen}, are not required here.


\subsection{\label{sec:theory:lambda}Estimation of the $\lambda$-line}

\begin{figure}[!t]
   \hspace*{-0.4cm}{\includegraphics{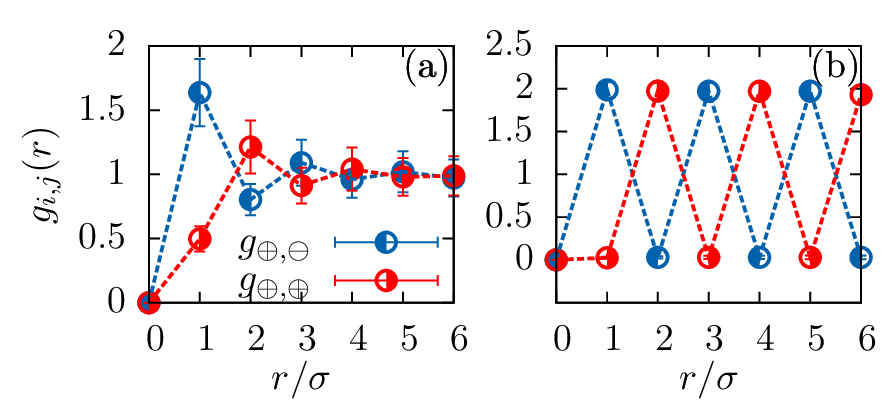}}
   \caption{
   Pair distribution functions for $L=12,\alpha=0,r_\text{cut}/\sigma
   =1$ using the $1$-norm ($p=q=1$) in Eq.~(\ref{eq:ham}) at temperature 
   $T^*=2$ and packing fraction $\eta=0.37$ (a) and $\eta=0.90$ (b).
   The red curves represent the pair distribution function for equally charged 
   ions, while the blue curves represent those of oppositely charged ions.
   Panel (a) shows the structure of the charge-disordered phase, which is 
   characterized by a rapid vanishing of the spatial correlations.
   Panel (b) displays the case of the charge-ordered phase, where correlations
   are long-ranged due to a shell-wise alternating distribution of cations and
   anions.
   }
   \label{fig:pdf_examples}
\end{figure}

The structure of the charge-ordered and charge-disordered phase can be observed
by analyzing the pair distribution functions $g_{i,j}(r)$
(see Eq.~(\ref{eq:pdf})). 
Figure~\ref{fig:pdf_examples} shows the respective structures of
both phases for the case $T^*=2,L=12,\alpha=0,r_\text{cut}/\sigma=1$
using the $1$-norm ($p=q=1$) in Eq.~(\ref{eq:ham}). 
For the charge-disordered phase, spatial correlations vanish rapidly, while for
the charge-ordered phase one observes long-ranged correlations due to a
shell-wise alternating assembly of cations and anions.

\begin{figure}[!t]
   \includegraphics{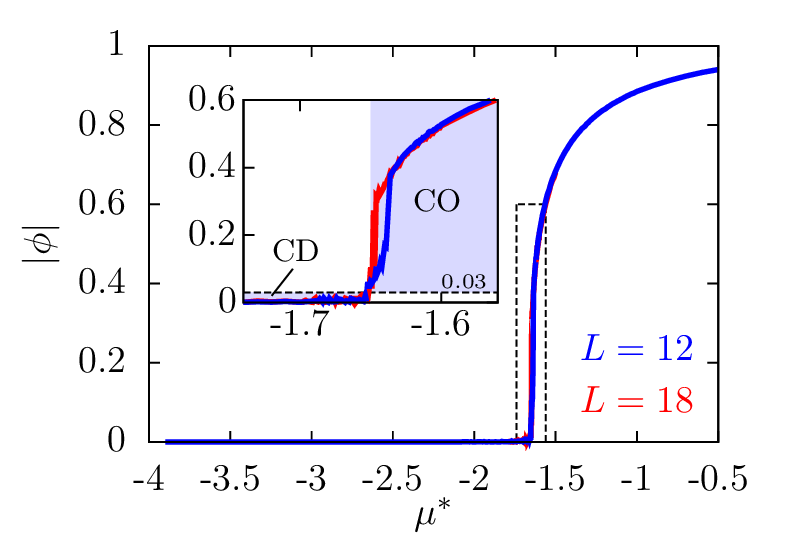}
   \caption{
   Order parameter $|\phi|$ as a function of the chemical potential
   $\mu^*$ for decay constant $\alpha=0$, cut-off radius 
   $r_\text{cut}/\sigma=1$ using the 1-norm ($p=q=1$) in Eq.~(\ref{eq:ham}),
   and temperature $T^*=1.67$ for box sizes $L\in\{12,18\}$.
   It signals the second-order phase transition between 
   the charge-disordered (CD) phase, where $|\phi|\approx0$, and the  
   charge-ordered (CO) phase, where $|\phi|\not\approx0$.
   In order to account for finite-size effects, the convention is adopted that
   the phase transition occurs at $|\phi|=0.03$, which is located at $\mu^*
   \approx-1.65$ here. 
   }
   \label{fig:deltag_example}
\end{figure}

In order to locate the charge-ordered/charge-disordered phase transition,
i.e., the $\lambda$-line, in the phase diagram, the so-called 
staggered order parameter
\begin{align}
   \phi := 
   \left\langle\frac{1}{L^3}\sum_{\mathbf{r}\in\mathcal{V}}
   (-1)^{||\mathbf{r}||_1}\zeta(\mathbf{r})\right\rangle
   \label{eq:opf}
\end{align}
is considered.
$|\phi|$ is positive in the charge-ordered phase and it vanishes in
the charge-disordered phase.
In order to account for finite-size effects, the convention is 
adopted that a thermodynamic state belongs to the charge-disordered phase iff
$|\phi|<0.03$.
Obviously, this convention is largely arbitrary, but it allows a sufficiently
precise localization of the $\lambda$-line.
Figure~\ref{fig:deltag_example} displays $|\phi|$ as a
function of the chemical potential $\mu^*$ for decay constant 
$\alpha=0$, cut-off radius $r_\text{cut}/\sigma=1$ and temperature 
$T^*=1.67$ for box sizes $L\in\{12,18\}$.
For this example, the second-order phase transition, which is 
expected to belong to the Ising universality class~\cite{Jain_Landau}, 
is located at $\mu^*\approx-1.65$.


\section{\label{sec:discussion}Results and Discussion}

\subsection{\label{sec:discussion:cutting_off}Sharp cut-off schemes}

In this subsection a sharp cutting off of the Coulomb potential, 
i.e., $\alpha=0$ in Eq.~(\ref{eq:phi_ij}), is considered with cut-off radii
$r_\text{cut}/\sigma\leq(L-1)/2$.
First, the discussion is concerned with the dependence
of the critical temperature $T_c^*$ on the cut-off radius
$r_\text{cut}$, which is obtained following the procedure described
in Subsec.~\ref{sec:theory:binodal}.

\begin{figure}[!t]
   \includegraphics{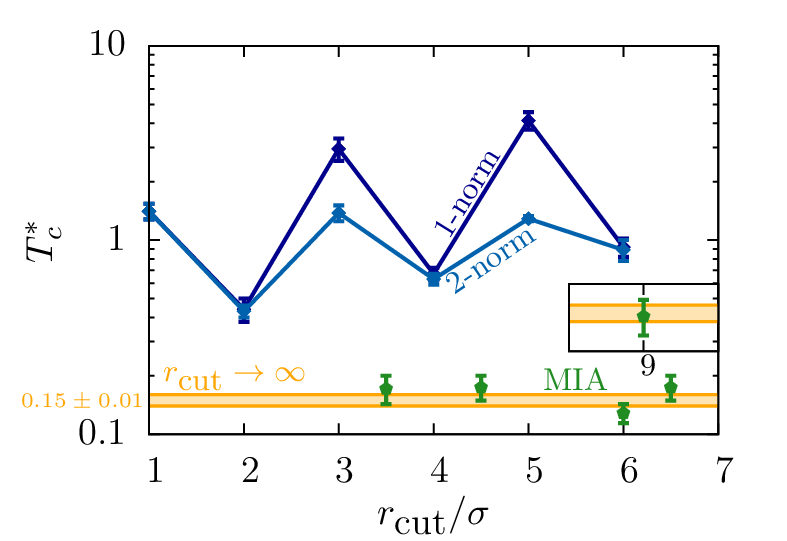}
   \caption{
   Critical temperature $T_c^*$ as a function of the cut-off radius
   $r_\text{cut}$ for sharp cut-off truncation schemes, 
   where the decay constant $\alpha=0$ is used in Eq.~(\ref{eq:phi_ij}).
   Data points labeled by '1-norm' and '2-norm' are obtained 
   by truncation of the Coulomb potential at distances
   $r_\text{cut}/\sigma\leq(L-1)/2$,
   where all distances are measured 
   in the respective norms, i.e., $p=q$
   (see Sec.~\ref{sec:discussion:cutting_off}).
   Moreover, the values within the minimum image approximation 
   ('MIA') are displayed, which corresponds to $p=2,q=\infty,
   r_\text{cut}/\sigma=(L-1)/2$ (see Sec.~\ref{sec:discussion:mia}).
   By comparison with the critical temperature $\Theta_c^*\in[0.14,0.16]$ for
   the long-ranged Coulomb potential ($r_\text{cut}=\infty$)
   one observes that the MIA results match the long-range value very
   well, whereas the 1-norm and the 2-norm results largely 
   overestimate the critical temperature.
   }
   \label{fig:Tc*}
\end{figure}

Figure~\ref{fig:Tc*} compares $T^*_c$ using the 1-norm ($p=q=1$) and 
the 2-norm ($p=q=2$) in Eq.~(\ref{eq:ham}) with the critical temperature 
$\Theta^*_c\in[0.14,0.16]$ of the long-ranged Coulomb system, i.e., for 
$r_\text{cut} = \infty$ \cite{2Panagiotopoulos_Kumar,Dickman_Stell,Ren_et_al}.
Distinct oscillations of $T_c^*$ for both 1-norm and 2-norm 
can be observed, where the amplitude of the latter is much smaller
than that of the former.
For both metrics, no signs of convergence of $T^*_c$ towards the value 
$\Theta^*_c$ of the long-ranged potential are observable within
the considered range of cut-off radii $r_\text{cut}$.
Moreover, the amplitude of the oscillations appears to
even increase for the 1-norm.
This odd-even-dependence likewise occurs within the structure for sufficiently
large packing fractions:
While for odd values of $r_\text{cut}/\sigma$ a charge-ordered phase is 
realized at large packing fractions, no charge-ordering is observed for even
values of $r_\text{cut}/\sigma$.

In order to understand the presence or absence of a charge-ordered
phase in the case of odd or even values of $r_\text{cut}/\sigma$, respectively,
consider the particular case of the 1-norm metric ($p=q=1$ in 
Eq.~(\ref{eq:ham})) and a perfectly charge-ordered configuration 
$\zeta_\text{CO}$.
The energy of such a configuration is given by (see Eq.~(\ref{eq:ham}))
\begin{align}
   \beta H[\zeta_\text{CO}]
   &=\frac{L^3}{2T^*}\sum_{i=1}^{r_{\text{cut}}/\sigma}
     \frac{(-1)^i}{i}V_\text{shell}(i\sigma)\notag\\
   &=\frac{L^3}{2T^*}\left(-6+\frac{18}{2}-\frac{38}{3}
                           +\frac{66}{4}-\frac{102}{5}+\cdots\right),
\label{eq:crystal_energy_rcut}
\end{align}
where $V_\text{shell}(i\sigma)=4i^2+2$ is the number of sites in the
$i$-th shell of 1-norm distance $i\sigma$ around a site.
For $r_\text{cut}/\sigma=1$ only nearest neighbors interact and
due to $\beta H[\zeta_\text{CO}]<0$ it is energetically favorable
for the ions to arrange alternatingly for sufficiently large $\eta$.
However, for $r_\text{cut}/\sigma=2$ the coions in the second shell 
overcompensate the contribution of the counterions in the
first shell, such that $\beta H[\zeta_\text{CO}]>0$, which renders the 
charge-ordered phase unfavorable as compared to the charge-disordered phase.
This pattern repeats upon increasing $r_\text{cut}$ with $\beta 
H[\zeta_\text{CO}]<0$ for odd and $\beta H[\zeta_\text{CO}]>0$ for even values
of $r_\text{cut}/\sigma$.
Since the size of the $i$-th shell grows $\propto i^2$ 
and the strength of the interaction with the central ion decreases
$\propto 1/i$, the total energy $\beta H[\zeta_\text{CO}]$
oscillates with an amplitude which increases $\propto i$.
This growing stability of the charge-ordered phase for odd values of
$r_\text{cut}/\sigma$ causes the increase in the critical temperature $T_c^*$
for odd-valued cut-off radii.
Furthermore, for all cases studied here,
it is found that the total energies within the charge-ordered phases
for odd-valued cut-off distances are much lower than for the long-ranged
Coulomb potential and therefore that the stability of the latter's 
charge-ordered phase is much weaker.
This leads to a significantly lower critical temperature compared to those,
obtained using the sharp cut-off scheme.
The absence of a charge-ordered phase in the case of even-valued cut-off radii
and the increasing deviation of the critical temperature $T^*_c$ from that of
the conventional LRPM with long-ranged Coulomb potential, $\Theta^*_c\in[0.14,
0.16]$ in the case of odd-valued cut-off radii leads to the conclusion of some
major defect of the sharp cut-off scheme.

Calculations using the 2-norm lead to qualitatively similar results.
The quantitative differences for $r_\text{cut}/\sigma>1$
to the case of a 1-norm metric can be 
attributed to the occurrence of non-integer-valued distances and to
the fact that the shells of constant distance from a central site
are spheres and not octahedrons.
Note, that for the case $r_\text{cut}/\sigma=1$ the distinction
between 1-norm and 2-norm is irrelevant.
Remarkably, the critical temperatures $T^*_c$ for 
odd values of $r_\text{cut}/\sigma$ are considerably smaller than 
those for the 1-norm metric, whereas for even 
values of $r_\text{cut}/\sigma$ the differences are very small.

\begin{figure}[!t]
   \includegraphics{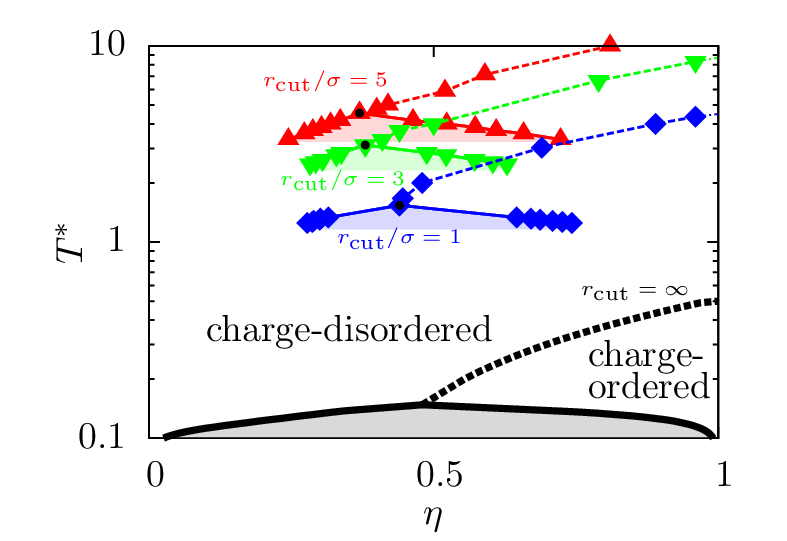}
   \caption{
   Phase diagrams for 1-norm metric ($p=q=1$ in Eq.~(\ref{eq:ham})), 
   decay constant $\alpha=0$, box size $L=12$,
   and cut-off radii $r_\text{cut}/\sigma\in
   \{1,3,5\}$ compared with that of the conventional long-ranged LRPM 
   ($r_\text{cut}=\infty$ \cite{Ren_et_al}).
   All these phase diagrams exhibit the same topology of a 
   charge-ordered and a charge-disordered phase separated by a first-order
   phase transition below and a continuous phase transition above a 
   tricritical point ($\bullet$).
   However, for the cases with finite values of the cut-off radius 
   $r_\text{cut}$, the tricritical temperature is about an order of magnitude
   too high and it increases upon increasing $r_\text{cut}$.
   }
   \label{fig:phasediagrams_cutting-off}
\end{figure}

Figure~\ref{fig:phasediagrams_cutting-off} displays the phase 
diagrams for cut-off radii $r_\text{cut}/\sigma\in\{1,3,5\}$ (1-norm, $p=q=1$), 
which have been determined using the methods presented in 
Secs.~\ref{sec:theory:binodal} and \ref{sec:theory:lambda}.
Close to the critical point, the binodals follow a straight line,
which corresponds to the critical exponent $\beta=1$.
Due to this observation and since the $\lambda$-line 
terminates at the critical point, this critical point 
actually is a \emph{tricritical} point (tagged by a black dot). 
The long-ranged LRPM qualitatively shows the same phase diagram, including 
tricriticality.
In accordance with the results of Fig.~\ref{fig:Tc*}, the tricritical point
moves upwards to higher temperatures for larger odd values
of $r_\text{cut}/\sigma$; however, the topology of the phase diagram
is not affected.
It has already been mentioned that for even values of the cut-off 
radius $r_\text{cut}/\sigma$ no charge-ordered phase is present so that a phase
separation occurs between a charge-disordered gas and a charge-disordered
liquid, and the phase diagram exhibits an ordinary critical point and no
$\lambda$-line.
Since the phase diagrams for even-valued cut-off radii are qualitatively
different from that of the long-ranged LRPM they are not shown in 
Fig.~\ref{fig:phasediagrams_cutting-off}.


\subsection{\label{sec:discussion:mia}Minimum image approximation (MIA)}

\begin{figure}[!t]
   \includegraphics{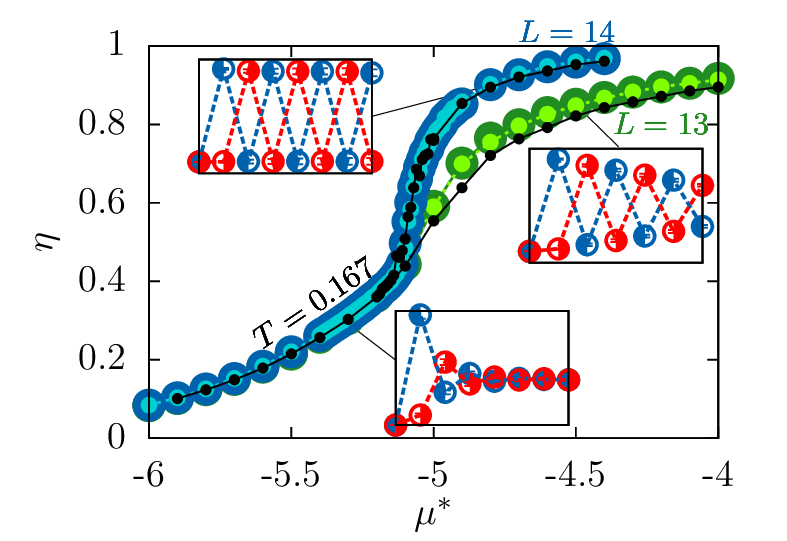}
   \caption{
   Comparison of the equations of state and of the pair
   distribution functions for odd and even box sizes $L$.
   The green and blue dots in the main plot correspond to results
   within MIA ($p=2,q=\infty,\alpha=0,r_\text{cut}/\sigma=(L-1)/2$
   in Eq.~(\ref{eq:phi_ij}))
   and the black dots are obtained by means of the Ewald method.
   The same odd-even-effect of $L$ in the high-density regime is
   encountered within both methods.
   Furthermore, the high-density phase displays decorrelations 
   in the case of odd values of $L$, whereas the charge ordering
   is long-ranged in the case of even values of $L$.
   The occurrence of decorrelations in the high-density phase for odd-valued
   $L$ can be understood in terms of equally charged ions at the rim
   of the simulation box, which interact with each other due to periodic
   boundary conditions.}
   \label{fig:mia_odd_even}
\end{figure}

When discussing the sharp cut-off schemes, i.e., with decay constant
$\alpha=0$ in Eq.~(\ref{eq:phi_ij}), in the previous 
Subsec.~\ref{sec:discussion:cutting_off}, cut-off radii $r_\text{cut}/\sigma\leq
(L-1)/2$ have been assumed throughout.
The phase diagram for odd values of the cut-off radius $r_\text{cut}/\sigma\leq
(L-1)/2$ turned out to be qualitatively identical to the phase diagram of the 
conventional long-ranged LRPM, but the tricritical point is located at a too 
high temperature $T^*_c$, which even increases upon increasing $r_\text{cut}$.
Here, another sharp cut-off scheme is discussed which, in the notation of
Eq.~(\ref{eq:phi_ij}), can be specified by $\alpha=0,q=\infty,
r_\text{cut}/\sigma = (L-1)/2$.
This cut-off scheme is equivalent to the well-known minimum image 
approximation (MIA) \cite{Metropolis_et_al,Card_Valleau}, which disregards all
contributions to the interaction energy, which do not correspond to
the minimum distance between two ions or their periodic images.
For a cubic simulation box, it can be interpreted such that only
those interactions within a cut-off distance $\|\vec{r}\|_\infty \leq
r_\text{cut}$ are taken into account, where the cut-off radius 
$r_\text{cut}$ has to be chosen such that the interaction range equals a cube
of volume $L^3$ with $L=2r_{\textnormal{cut}}/\sigma+1$.
Note, that the distances, which determine interaction potential, are measured
in the $p$-norm (see Eq.~(\ref{eq:phi_ij})), which is chosen as $p\in\{1,2\}$.

Figure~\ref{fig:Tc*} also displays the critical temperatures $T^*_c$ 
within MIA for $p=2$, which are in excellent agreement with the
critical temperature $\Theta^*_c\in[0.14,0.16]$ of the long-ranged LRPM.
Moreover, the MIA results in Fig.~\ref{fig:Tc*} can be subdivided
into integer and half-integer values of 
$r_\text{cut}/\sigma$, which corresponds respectively to odd and even values
of the box size $L=2r_\text{cut}/\sigma+1$.
A close look at Fig.~\ref{fig:Tc*} reveals that the values of $T^*_c$
within MIA for odd values of $L$ are slightly below 
those for even values of $L$, which indicates
a certain odd-even effect of $L$ on $T^*_c$.
Whereas the odd-even effect of $L$ on $T^*_c$ is rather weak, 
Fig.~\ref{fig:mia_odd_even} shows for the examples $L=13$ and $L=14$ that there
is a significant odd-even effect of $L$ on the equation of state and on the
structure at large packing fractions.
The reason for mentioning the odd-even effects of $L$ within MIA here is that
exactly the same odd-even effects of $L$ occur for the long-ranged LRPM using
the full Ewald method, which are shown in Fig.~\ref{fig:mia_odd_even}, too.
In other words, MIA is such a good approximation of the full Ewald method that
it exhibits even the same odd-even effects.

The insets in Fig.~\ref{fig:mia_odd_even}, which show the pair
distribution functions at the indicated thermodynamic states, 
reveal a distinct charge-ordering behavior in the high-density 
regime for even box sizes $L$, whereas for odd $L$ a decaying 
oscillatory behavior of the pair distribution functions can be observed.
Although the latter behavior does not correspond to a 
genuine charge-ordered phase, it is clearly different from the structure in 
the dilute regime, where correlations decay rapidly.
What is the reason for this qualitatively different high-density 
structure for even and odd box sizes $L$?
In the limit of perfect charge-ordering at sufficiently high densities,
the periodic boundary conditions lead to destabilizing contributions 
to the interaction energy for odd values of $L$, since in
this case there are equally charged ions located at the rim of the
simulation box which, due to the periodicity, are nearest neighbors.
This effect does not occur for even box sizes $L$. 
Since this is obviously an effect related to the `surface' of the 
simulation box, it decreases with increasing box size $L$.
The same odd-even dependence of the equation of state and of the structure is 
observable within the long-ranged LRPM.
As a consequence of the above argumentation, the results for even box sizes 
$L$, whether using MIA or the full Ewald method, can be expected to
be more accurate, than for odd box sizes $L$.

It is striking that isotherms within MIA sample the long-range limit
$r_\text{cut}\to\infty$ very well for dilute as well as for dense
systems.
This indicates that the Hamiltonian of an ionic 
fluid with long-ranged Coulomb interaction can be reliably
approximated by Eq.~(\ref{eq:ham}) with $p=2,q=\infty,\alpha=0,
r_\text{cut}/\sigma=(L-1)/2$ in Eq.~(\ref{eq:phi_ij}).
In fact, for $L=14$, there is almost no difference between MIA and
the full Ewald method with respect to the equations of state (see 
Fig.~\ref{fig:mia_odd_even}).   

This quantitative agreement can be understood as follows: 
The contribution to the energy due to the long-ranged Coulomb potential beyond
MIA involves the electrostatic interaction of the simulation box with its
images, which, due to the charge neutrality of the simulation box, decays at
least as a dipole-dipole interaction, i.e., $\propto 1/r^3$.
However, numerical calculations reveal that the dipole-dipole contribution
vanishes and that the decay is actually $\propto 1/r^4$, which renders the 
long range contributions beyond MIA absolutely convergent and, for sufficiently
large $L$, small.
The same argument does not apply to the sharp cut-off potentials of 
Subsec.~\ref{sec:discussion:cutting_off}, since no charge neutrality inside
spheres of radius $r_\text{cut}$ within the underlying norm is guaranteed so 
that the long range contributions beyond those due to the sharp cut-off 
potential are due to an effective monopole-monopole interaction, i.e., 
$\propto 1/r$, and, hence, typically not small.

\begin{figure}[!t]
   \includegraphics{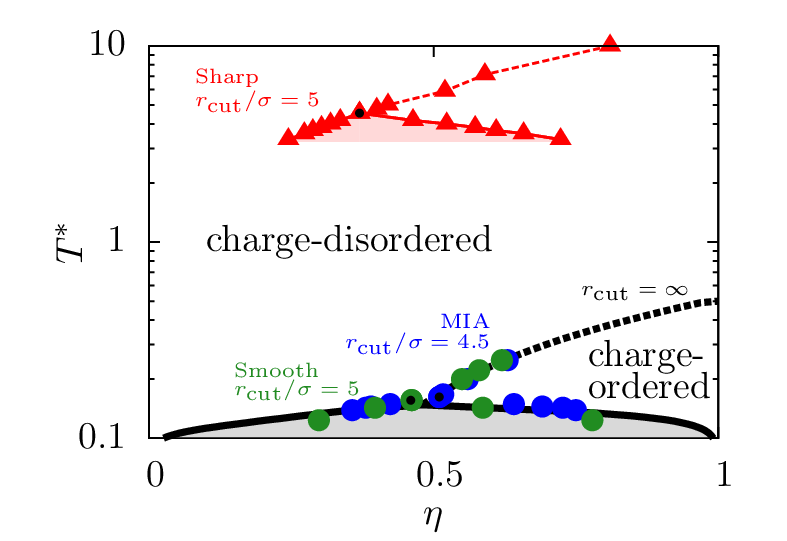}
   \caption{
   Comparison of the phase diagrams of the LRPM within the sharp 
   cut-off scheme $p=q=1,\alpha\sigma=0,r_\text{cut}/\sigma=5\leq(L-1)/2$ in 
   Eq.~(\ref{eq:phi_ij}), the MIA ($p=2,q=\infty,\alpha\sigma=0, r_\text{cut}/\sigma
   =4.5=(L-1)/2$), the smooth cut-off scheme
   ($p=q=2, \alpha\sigma=0.8,r_\text{cut}/\sigma=5\leq(L-1)/2$),
   and the long-ranged Coulombic interaction ($\alpha\sigma=0,
   r_\text{cut}=\infty$).
   The agreement of the MIA and the smooth cut-off scheme
   with the case of the long-ranged interaction is 
   excellent, whereas there are large quantitative deviations of the sharp 
   cut-off scheme with $r_\text{cut}/\sigma\leq(L-1)/2$.
   }
   \label{fig:phasediagrams_mia}
\end{figure}

Figure~\ref{fig:phasediagrams_mia} displays the phase diagram within
MIA for $r_\text{cut}/\sigma=4.5$; those for $r_\text{cut}/\sigma\in
\{6,6.5,9\}$ obtained similarly are not shown here.
For all considered cases of $r_\text{cut}$, the phase diagrams are 
qualitatively and quantitatively in accordance with those obtained by
means of the corresponding full Ewald method.
The important point is that the MIA can be interpreted as
a sharp cut-off truncation scheme of the Coulomb potential.
The results discussed above indicate that the long-range
character of the electrostatic interaction may not be necessary for
the thermal and structural properties of ionic fluids, 
because there is the possibility that the same properties can be 
generated by suitable short-ranged interactions.
A comparison of the MIA, where $r_\text{cut}/\sigma=(L-1)/2$, with the
sharp cut-off schemes of the previous Subsec.~\ref{sec:discussion:cutting_off},
where $r_\text{cut}/\sigma\leq(L-1)/2$, reveals that the ability of a sharp 
cut-off scheme to mimic the properties of the long-ranged LRPM depends 
delicately on the relation between the cut-off radius $r_\text{cut}$ and the
size of the simulation box $L$.
However, the simulation box $L$ is not a physical but rather a technical 
parameter, and the properties of the LRPM within reasonable sharp cut-off 
schemes Eq.~(\ref{eq:phi_ij}) with $\alpha=0$ should be independent of (large
values of) $L$.
Consequently, there are no sharp cut-off schemes which quantitatively reproduce
the thermal and structural properties of the long-ranged LRPM and which, at
the same time, correspond to physically acceptable underlying short-ranged 
interactions.


\subsection{\label{sec:discussion:short_range}Smooth cut-off potentials}

It turned out in the previous Subsec.~\ref{sec:discussion:mia} that
there are sharp cut-off schemes, i.e., with $\alpha=0$ in Eq.~(\ref{eq:phi_ij}),
which, although based on a short-ranged potential Eq.~(\ref{eq:phi_ij}), lead
to quantitatively identical thermal and structural properties as the 
conventional LRPM with the long-raged Coulomb potential.
However, these sharp cut-off schemes do not correspond to physical systems,
because one has to tune the cut-off radius $r_\text{cut}$ according to the
unphysical simulation box size $L$, whereas for physically acceptable
short-ranged potentials $r_\text{cut}$ should be independent of $L$.
Here, potentials Eq.~(\ref{eq:phi_ij}) with $\alpha>0$ are considered, which
correspond to a smooth cutting off of the Coulomb potential on the length scale
$1/\alpha$.
Note, that this type of functions occurs within the Ewald method 
as a result of splitting the total electrostatic potential into 
a short-ranged contribution, which leads to a sum in real space, and
a long-ranged contribution, which leads to a sum in reciprocal 
space. 
Within the Ewald method, $\alpha$ is adjusted such that, on the one hand, the 
error due to the unavoidable truncation of the sums in 
real and reciprocal space are sufficiently small and that, on the other
hand, the computational effort is acceptable.
However, here $\alpha$ controls the decay length of a genuine short-ranged 
interaction, which can take, in principle, any (positive) value. 
For $\alpha\to0$ the sharp cut-off truncation schemes discussed in 
Sec.~\ref{sec:discussion:cutting_off} are obtained.

\begin{figure}[!t]
   \includegraphics{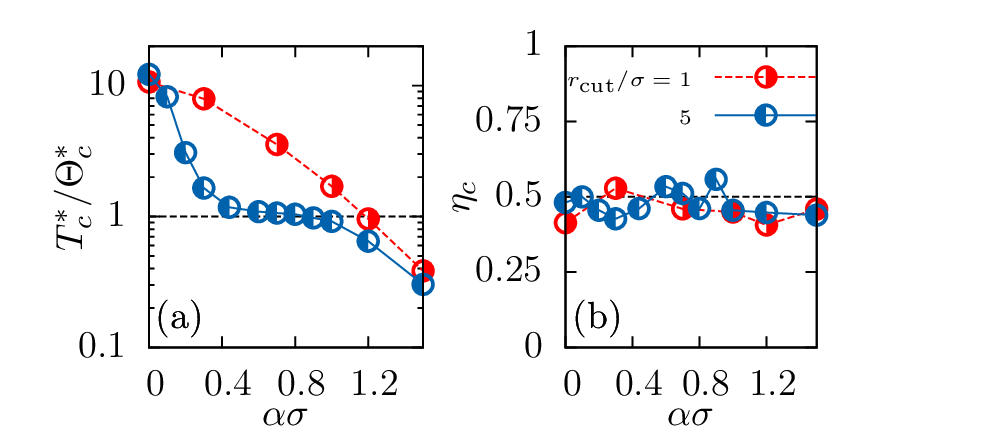}
   \caption{
   Critical temperature $T^*_c$ and critical packing fraction 
   $\eta_c$ of the LRPM with smooth cut-off potentials Eq.~(\ref{eq:phi_ij})
   with $p=q=2,L=12,r_\text{cut}/\sigma\in\{1,5\}$ as functions of the decay
   constant $\alpha\sigma$.
   For $r_\text{cut}/\sigma=5$ the critical temperature $T^*_c$ exhibits a
   plateau around $\alpha\sigma\approx0.8$ which is close to the critical
   temperature $\Theta_c^*$ of the long-ranged Coulombic system.
   As in Figs.~\ref{fig:Tc*}, \ref{fig:phasediagrams_cutting-off},
   and \ref{fig:phasediagrams_mia}, the critical temperature $T^*_c$ within
   sharp cut-off schemes ($\alpha\sigma=0$) is an order of magnitude to high.
   The critical packing fraction $\eta_c\approx0.5$ is rather independent of
   the decay constant $\alpha$ and the cut-off radius $r_\text{cut}$.
   }
   \label{fig:alpha_Tc_etac}
\end{figure}

Figure~\ref{fig:alpha_Tc_etac} displays the critical temperature 
$T^*_c$ and the critical packing fraction $\eta_c$ of the LRPM with smooth 
cut-off potentials Eq.~(\ref{eq:phi_ij}) with $p=q=2,r_\text{cut}/\sigma\in
\{1,5\}$ as functions of the decay constant $\alpha\sigma$.
For $r_\text{cut}/\sigma=5$ a plateau of $T_c^*$ is present in the range
$\alpha\sigma\in[0.4,1.0]$, where the critical temperature $T^*_c$ is
quantitatively equivalent to the critical temperature $\Theta_c^*$ of the 
long-ranged LRPM (see Fig.~\ref{fig:alpha_Tc_etac}(a)).
For larger values of $\alpha\sigma$ the decay length of the 
interaction Eq.~(\ref{eq:phi_ij}) becomes too short, so that even 
nearest neighbors barely interact with each other.
Hence, for $\alpha\sigma\to\infty$, one obtains the ideal gas limit and 
consequently $T_c^*\to0$.
For $r_\text{cut}/\sigma=1$ all contributions beyond the nearest neighbors
are neglected, and no plateau is observable at all.
However, the critical packing fraction $\eta_c\approx0.5$ appears to
be independent of the decay constant $\alpha$ and the cut-off radius 
$r_\text{cut}$.
In addition to the coincidence of the critical points for the appropriate
choice of $\alpha\sigma$, a charge-ordered phase at large
packing fractions and a charge-disordered phase at small packing fractions
can be found.
Ultimately, the phase diagrams of the smooth cut-off scheme
are in quantitative agreement with the long-ranged limit, too
(see Fig.~\ref{fig:phasediagrams_mia}).

\begin{figure}[!t]
   \includegraphics{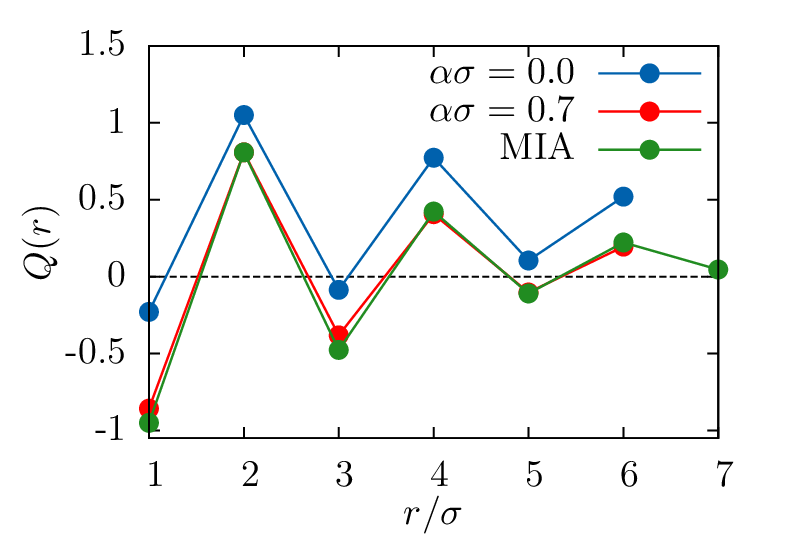}
   \caption{
   Accumulated charge $Q(r)$ (see Eq.~(\ref{eq:StillingerLovettFirstMoment}))
   up to a distance $r$ in 1-norm around a positively charged central ion.
   The Stillinger-Lovett sum rule \cite{Stillinger_Lovett,Lovett_Stillinger,
   Stillinger_Lovett2,Hansen2006} requires $Q(r)$ to vanish in the limit of
   large distances $r\to\infty$.
   This condition appears to be fulfilled within the
   charge-disordered phase (here at packing fraction $\eta\approx0.3$) for the
   case of the smooth cut-off potential $\alpha\sigma=0.7$
   ($p=q=2,r_\text{cut}/\sigma=5,L=12$)
   as well as for the MIA
   ($\alpha\sigma=0,p=2,q=\infty,r_\text{cut}/\sigma=6.5=(L-1)/2,L=14$),
   but not for the sharp cut-off potential $\alpha\sigma=0$
   ($p=q=2,r_\text{cut}/\sigma=1,L=12$).
   }
   \label{fig:perfect_screening}
\end{figure}

The long-range character of the Coulomb interaction is well-known to
generate the so-called Stillinger-Lovett perfect screening property
of ionic fluids \cite{Stillinger_Lovett,Lovett_Stillinger,Stillinger_Lovett2,
Hansen2006,Mitchell_et_al,Martin_Gruber}. 
In the present context of an LRPM it corresponds to the accumulated
charge
\begin{align}
    Q(r) & :=1 + \frac{\eta}{2}
    \sum_{\substack{\vec{r}'\in\mathcal{V}\\1\leq\|\vec{r}'\|_1\leq r}}
    \left[g_{\oplus,\oplus}(\mathbf{r}') -
          g_{\oplus,\ominus}(\mathbf{r}')\right]
   \label{eq:StillingerLovettFirstMoment}
\end{align}
up to a 1-norm distance $r$ around a positively charged central ion to vanish
$Q(r)\to0$ in the limit $r\to\infty$.
Figure~\ref{fig:perfect_screening} compares the accumulated charge $Q(r)$ for
the sharp cut-off potential with $\alpha\sigma=0,r_\text{cut}/\sigma=1,p=q=2$,
the MIA with $\alpha\sigma=0,r_\text{cut}/\sigma=6.5,p=2,q=\infty$
and the smooth cut-off potential $\alpha\sigma=0.7,r_\text{cut}/\sigma=5,p=q=2$
within the charge-disordered phase at packing fraction $\eta\approx0.3$.
For the sharp cut-off potential, $Q(r)$ appears to converge towards a 
non-vanishing (positive) value for large radii $r$, which corresponds to an
imperfect screening of the positive central charge.
However, perfect screening occurs within the MIA and the smooth cut-off 
potential. 
This is an interesting finding since, in contrast to the long-range Coulomb
potential, the perfect screening property is not necessarily fulfilled in
systems with short-ranged interactions.

It can be shown for smoothly cut-off potentials Eq.~(\ref{eq:phi_ij}) with 
$r_\text{cut}\alpha\gg1$ that the charge-charge pair correlation function
$h_{zz}(r):=2\left(g_{\oplus,\oplus}(r)-g_{\oplus,\ominus}(r)\right)$, which
is related to the accumulated charge $Q(r)$ in Eq.~(\ref{eq:StillingerLovettFirstMoment}),
decays asymptotically on the length scale of the Debye length $1/\kappa$ with
$(\kappa\sigma)^2=4\pi\eta/T^*$ iff $\kappa\gg2\alpha$.
Since $\alpha>0$ is a constant, which is chosen to match the phase diagram of
the long-ranged Coulomb interaction (see Figs.~\ref{fig:phasediagrams_mia} and
\ref{fig:alpha_Tc_etac}) and which does not depend on the thermodynamic state 
$(\eta,T^*)$, the condition $\kappa\gg2\alpha>0$ will not be fulfilled at very 
low packing fractions $\eta$ and/or high temperatures $T^*$, i.e., the 
Debye-H\"{u}ckel limit is not recovered within smooth cut-off schemes.
However, we do not consider this a major defect for two reasons:
On the one hand, the remaining parts of the phase diagram outside the region of
the Debye-H\"{u}ckel limit, particularly in the range of high densities, e.g.,
close to the critical point ($\eta_c\approx0.5$), are reproduced quantitatively
(see Figs.~\ref{fig:phasediagrams_mia} and \ref{fig:alpha_Tc_etac}).
On the other hand, the crossover, where the condition $\kappa\gg2\alpha$ begins
to be violated, can be shifted to arbitrarily small values of $\kappa$ by
decreasing $\alpha$, which can be achieved by accordingly increasing $r_\text{cut}$
(see Fig.~\ref{fig:alpha_Tc_etac}(a)).

Concerning the phase behavior and the structure, the LRPM
with the smooth cut-off potentials considered here is qualitatively equivalent
to that with long-ranged Coulomb interactions.
Moreover, by choosing an appropriate decay constant $\alpha$, the
short-ranged smooth cut-off potential LRPM becomes even 
quantitatively equivalent to the conventional long-ranged LRPM.
The essential difference to the sharp cut-off schemes considered in 
Subsecs.~\ref{sec:discussion:cutting_off} and \ref{sec:discussion:mia} is that
these statements are independent of the choice of the actual (large) value of
the simulation box $L$, i.e., the smooth cut-off potentials discussed here
are candidates of physically meaningful short-ranged interactions.
Consequently, the long-range character of the electrostatic potentials is of
minor importance for the thermal and structural properties of ionic fluids,
since, as has been demonstrated here, short-ranged interaction
potentials do exist, which lead to the same thermal and structural properties.


\section{\label{sec:conclusions}Conclusions and Summary}

In order to elucidate the role of the long-range character of the 
Coulomb potential for the thermodynamic and the structural 
properties of ionic fluids, the lattice restricted primitive model
(LRPM) with various truncated Coulomb-like interaction potentials 
Eq.~(\ref{eq:phi_ij}) has been studied in this work by means of grandcanonical
Monte Carlo simulations.

The simplest approach of sharply truncating the Coulomb
potential at a certain cut-off radius $r_\text{cut}$ 
(Sec.~\ref{sec:discussion:cutting_off}) turned out to be not 
appropriate to reproduce the properties of the long-ranged LRPM, either 
qualitatively or quantitatively.
When measuring distances within the 1-norm, e.g., no charge-ordered phase 
occurs for even-valued cut-off radii $r_\text{cut}/\sigma$, i.e., the topology
of phase diagram differs from that of the conventional long-ranged LRPM.
On the other hand, odd-valued cut-off radii lead to qualitatively the same 
phase diagrams as the conventional long-ranged LRPM 
(Fig.~\ref{fig:phasediagrams_cutting-off}), but there is a large quantitative
difference, e.g., of the value of the critical temperature $T^*_c$
(Fig.~\ref{fig:Tc*}).

In contrast, the well-known minimum image approximation (MIA), which
may be viewed as a sharp cut-off scheme with the cut-off radius $r_\text{cut}$
being related to the side length of the simulation box $L$, leads to 
quantitative agreement of the phase behavior and the structure of the LRPM with
that of the long-ranged LRPM (Fig.~\ref{fig:phasediagrams_mia}).

However, due to the relation of the cut-off radius with the unphysical
parameter $L$, the MIA does not correspond to a physically acceptable, i.e.,
$L$-independent, short-ranged potential.
Quantitative agreement of the phase behavior and the structure of short-ranged
and long-ranged LRPMs is achieved by smoothly cutting off the Coulomb 
interaction, which is expressed by a decay constant $\alpha$, corresponding to
a decay length of $1/\alpha$ (Fig.~\ref{fig:alpha_Tc_etac}).
Interestingly, the LRPM with the smoothly cut off Coulomb potential possesses,
although it is based on short-ranged potentials, the Stillinger-Lovett perfect
screening property (Fig.~\ref{fig:perfect_screening}).

By showing that the characteristic phase behavior and structure of 
ionic fluids can also be found in systems with short-ranged potentials, one can
conclude that the decisive property of the electrostatic potential in ionic
fluids is not the long range but rather the valency dependence.
As this conclusion is drawn from a study of the LRPM, the natural question 
arises whether it holds for off-lattice models, too.
This task is left for future investigations.




\bibliography{literature}


\end{document}